\documentclass[showpacs,aps,graphicx,twocolumn]{revtex4}%,,preprint
\usepackage{amssymb}
\usepackage{txfonts}%and
\usepackage{graphicx}

\begin{document}

\title{Optimal multipartite entanglement concentration of electron-spin states  based on charge detection and projection measurements}

\author{B. C. Ren$^1$, T. J. Wang$^{1,2}$, M. Hua$^1$, F. F. Du$^1$, and F. G. Deng$^1$\footnote{Corresponding author. Email address:
fgdeng@bnu.edu.cn} }
\address{ $^1$ Department of Physics, Applied Optics Beijing Area Major Laboratory, Beijing Normal University, Beijing 100875, China\\
$^2$ Department of Physics, Tsinghua University, Beijing 100084,
China }
\date{\today }

\date{\today }

\begin{abstract}
We propose an optimal entanglement concentration protocol (ECP) for
nonlocal $N$-electron systems in a partially entangled pure state,
resorting to charge detection and the projection measurement on an
additional electron. For each nonlocal $N$-electron system, one
party in quantum communication, say Alice first entangles it with an
additional electron, and then she projects the additional electron
into an orthogonal basis for dividing the  $N$-electron systems into
two groups. In the first group, the $N$ parties  obtain a subset of
$N$-electron systems in a maximally entangled state directly. In the
second group, they obtain some less-entangled $N$-electron systems
which are the resource for the entanglement concentration in the
next round. By iterating the entanglement concentration process
several times, the present ECP has the maximal success probability,
the theoretical limit of an ECP as it just equals to the
entanglement of the partially entangled state, far higher than
others, without resorting to a collective unitary evolution.
\end{abstract}

 \maketitle

%Entanglement production and manipulation - Quantum error correction
%and other methods for protection against decoherence -

%Decoherence; open systems; quantum statistical methods -

%

\section{Introduction}

The principles in quantum mechanics provide some novel ways for
secure communication. Since Bennett  et al. \cite{bb84} published
the original quantum key distribution (QKD) protocol in 1984,
quantum communication has attracted a lot of attention. For example,
Ekert \cite{Ekert91} proposed a QKD protocol based on two-photon
entanglement in 1991.  Subsequently,  Bennett,  Brassard, and Mermin
\cite{BBM92} simplified its process for eavesdropping check. In
1992, Bennett and Wiesner proposed a quantum dense coding protocol
between two parties \cite{densecoding} and it was generalized to $N$
parties with arbitrary $d$-dimensional quantum systems  by Liu
\emph{et al.} \cite{super2} in 2002. In 1993, Bennett \emph{et al.}
proposed a quantum teleportation protocol \cite{teleportation} for
transmitting an unknown single-qubit state, without moving the qubit
itself by setting up a quantum channel with a photon pair in a
maximally entangled state. In 1999, Hillery, Bu\v{z}ek, and
Berthiaume \cite{QSS} presented an interesting quantum secret
sharing protocol based on multipartite photon systems in a maximally
entangled state. Subsequently, it is generalized to the case with
two-photon entangled channels \cite{QSS2}, arbitrary number of
agents \cite{QSS3},  and that to sharing an unknown quantum state
\cite{QSTS,QSTS2,QSTS3,QSTS4} with a quantum channel in a
multipartite maximally entangled state. In 2002, Long and Liu
\cite{LongLiu} proposed the first quantum secure direct
communication (QSDC) protocol with a block of two-photon systems in
Bell states. It was detailed in the two-step QSDC protocol
\cite{two-step} and was generalized to the case with sing photons
\cite{QOTP} and  high-dimensional entangled quantum systems
\cite{highdimension,wangtjcpl}.

Although there are some interesting quantum communication protocols,
including those based on single photons \cite{bb84} or weak pulses
\cite{faint1,faint2,faint3,faint4}, quantum repeaters are required
in long-distance quantum communication \cite{Duan} because quantum
signals can only be transmitted over an optical fiber or a free
space not more than several hundreds kilometers with current
technology. In a quantum repeater, entanglement is used to connect
the two neighboring nodes. That is, the distribution of entanglement
between two nodes is necessary for a quantum repeater. Moreover, the
parties should store the quantum state for linking the other nodes
in quantum communication network. In a practical transmission of a
photon system, it will inevitably interact with its environment.
That is, it will suffer from the channel noise. In the storage of
the entangled quantum state, the decoherence will also decrease the
entanglement of the system. The decoherence of the entangled quantum
system will decrease the security of QKD protocol and even make a
quantum teleporation and a quantum dense coding protocol fail.

%to purify two-photon systems in a Werner state

Entanglement purification is used to extract a subset of
high-fidelity entangled systems from a set of less-entangled systems
in a mixed state. Since Bennett  et al. \cite{Bennett1} proposed the
first entanglement purification protocol (EPP) in 1996, many works
were focused on it and many important EPPs have been proposed,
resorting to different physical systems
\cite{Bennett1,Deutsch,Pan1,Simon,shengpra,shengpratwostep,shengpraonestep,lixhonestep,dengonestep,dengEMEPP,wangcqic,wangcoe,shengepjd,shengepppla}.
Compared with an EPP,  an entanglement concentration  protocol (ECP)
is usually more efficient for the two parties in quantum
communication, say Alice and Bob, to distill some maximally
entangled systems from an ensemble in a less-entangled pure state.
Up to now, there are some interesting ECPs
\cite{Bennett2,Yamamoto,zhao1,shengpraecp,shengsingle,swapping1,swapping2,shengecp}.
For example, Bennett \emph{et al.} \cite{Bennett2} proposed the
first ECP for two-photon systems in 1996, called it the Schmidt
projection method. In 1999, Bose et al.  \cite{swapping1} proposed
another interesting ECP based on entanglement swapping of two photon
pairs in a partially entangled pure state. Subsequently, Shi  et al.
\cite{swapping2} presented a different ECP based on entanglement
swapping and a collective unitary evolution.  Both these ECPs
\cite{swapping1,swapping2} require that Alice and Bob know the
information about the less-entangled pure state $\alpha \vert
H\rangle_{A}\vert H\rangle_{B} + \beta \vert V\rangle_{A}\vert
V\rangle_{B}$. Here $|H\rangle$ and $|V\rangle$ represent the
horizontal and the vertical polarizations of photons. In 2001,
Yamamoto \emph{et al.} \cite{Yamamoto} and Zhao et al.  \cite{zhao1}
proposed an ECP for photon pairs based on linear optical elements
independently. In 2008, Sheng, Deng, and Zhou \cite{shengpraecp}
proposed an ECP for photon systems based on cross-Kerr
nonlinearities. By iteration of the entanglement concentration
process, it has a far higher efficiency and yield than those in
Refs. \cite{Yamamoto,zhao1}. Both these ECPs
\cite{Yamamoto,zhao1,shengpraecp}  do not require that Alice and Bob
know the parameters $\alpha$ and $\beta$. Certainly, they has a
lower efficiency than those in Refs.\cite{swapping1,swapping2}. In
2010, the first single-photon ECP \cite{shengsingle} was discussed
with cross-Kerr nonlinearity. In 2012, Sheng  et al.
\cite{shengecp} proposed an interesting ECP for for partially
entangled photon pairs assisted by single photons.

An electron-spin system is an interesting qubit in quantum
computation and quantum communication. For example, Beenakker  et
al.  \cite{Beenakker} exploited the charge detection \cite{cd} to
construct a CNOT gate  based on both the charge and the spin degrees
of freedom of electrons in 2004. In 2007, Ionicioiu \cite{paritybox}
used charge detection to complete the generation of the entangled
spins. In 2006, Zhang, Feng, and Gao \cite{cluster} presented a
scheme for the multipartite entanglement analyzer. In 2005, Feng,
Kwek, and Oh \cite{feng} proposed an EPP for two-electron systems
based on charge detection. In 2011, an EPP for multi-electron
systems \cite{shengpla} was proposed. Moreover, the entanglement
concentration for multi-electron systems was discussed by Sheng
\emph{et al.} \cite{shengplaec}  based on charge detection and an
ECP for two-electron systems was proposed by Wang \emph{et al.}
\cite{wangcpra} based on quantum dots in micro-wave cavities.

In this paper, we proposed an optimal ECP for nonlocal $N$-electron
systems in a partially entangled pure state, resorting to the
projection measurement on an additional electron. In the present
ECP, Alice first entangles each nonlocal $N$-electron system with an
additional electron by performing a parity-check measurement on her
electron $A$ and an additional electron $a$, and then she projects
the electron $a$ into an orthogonal basis $\{\vert\varphi\rangle,
\vert\varphi^\bot\rangle\}$ for dividing the  $N$-electron systems
into two groups. In the first group, the $N$ parties  in quantum
communication  obtain the $N$-electron systems in a maximally
entangled state directly. In the second group, they  obtain some
$N$-electron systems in another partially entangled state with less
entanglement, which are the resource for entanglement concentration
in the next round. By iterating the process several times, the
present ECP has an optimal success probability, the theoretical
limit as it is just the entanglement of the partially entangled
state $E$, twice of those based entanglement swapping and a
collective unitary evolution \cite{swapping1,swapping2}, far higher
than other typical ECPs
\cite{Yamamoto,zhao1,shengpraecp,shengplaec,wangcpra}. Moreover, it
does not require a collective unitary evolution, which decreases the
difficulty of its implementation.

\section{Optimal multipartite entanglement concentration of three-electron spin states}

Before we describe the principle of our ECP for  nonlocal
$N$-electron systems, we first introduce the principle of the
parity-check gate (PCG) for two electrons based on their charge
detection, similar to that in Ref.\cite{Beenakker}, shown in Fig.1.
The charge detector (C) can distinguish the occupation number one
from the occupation numbers 0 and 2, but it cannot distinguish the
electron numbers between 0 and 2. That is, it can distinguish the
occupation number even or odd, which means that this device can
distinguish the even parity states $\vert\uparrow\rangle_{A_1}\vert
\uparrow\rangle_{A_2}$ and
$\vert\downarrow\rangle_{A_1}\vert\downarrow\rangle_{A_2}$ from the
odd parity states $\vert\uparrow\rangle_{A_1}\vert
\downarrow\rangle_{A_2}$ and $\vert\downarrow\rangle_{A_1}\vert
\uparrow\rangle_{A_2}$. In detail, as the spin polarizing beam
splitter (PBS: 50/50) can transmit an electron in the spin-up state
$\vert\uparrow\rangle$ and reflect an electron in the spin-down
state $\vert\downarrow\rangle$, one can see that the states
$|\uparrow\uparrow\rangle$ and $\vert\downarrow\downarrow\rangle$
will lead the charge detection to have the charge occupation number
$C=1$ as each electron passes through a different path after the
first PBS. The states $\vert\uparrow\downarrow\rangle$ and
$\vert\downarrow\uparrow\rangle$ will lead the charge detection to
$C=0$ and $C=2$, respectively. The charge detection cannot
distinguish 0 and 2, and  it will show the same result, i.e., $C=0$
for simplicity. The states $\vert\uparrow\uparrow\rangle$ and
$\vert\downarrow\downarrow\rangle$ can be distinguished from
$\vert\uparrow\downarrow\rangle$ and
$\vert\downarrow\uparrow\rangle$ by the different outcomes of the
charge detection. So this device can be used to accomplish a parity
check on a two-electron system, without destroying it. After the
second PBS (PBS$_2$), the states and the positions of the two
electrons are recovered when the electrons emit from the outputs
$A_3$ and $A_4$, respectively. That is, the charge detection $C$ is
a nondestructive quantum nondemolition detection on the electron
spins \cite{Beenakker}.

\begin{figure}[!h]%[tpb]
\begin{center}
\includegraphics[width=7.6cm,angle=0]{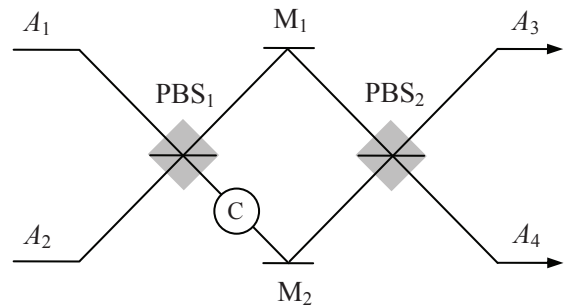}
\caption{The principle of our parity-check gate (PCG). Here PBS
represents a 50:50 spin polarizing beam splitter which is used to
transmit an electron in the spin-up state $\vert\uparrow\rangle$ and
reflect an electron in the spin-down state $\vert\downarrow\rangle$,
respectively. C represents a charge detection, which can distinguish
the occupation number one from the occupation number 0 and 2, but it
cannot distinguish the electron numbers between 0 and 2. $M_1$ and
$M_2$ are two mirrors. This setup can distinguish the even parity
states $\vert \uparrow\rangle_a\vert \uparrow\rangle_b$ and $\vert
\downarrow\rangle_a\vert \downarrow\rangle_b$ from the odd parity
states $\vert \uparrow\rangle_a\vert \downarrow\rangle_b$ and
$\vert\downarrow\rangle_a\vert  \uparrow\rangle_b$, without
destroying them.} \label{fig1_QND}
\end{center}
\end{figure}

\begin{figure}[!h]%[tpb]
\begin{center}
\includegraphics[width=7.2cm,angle=0]{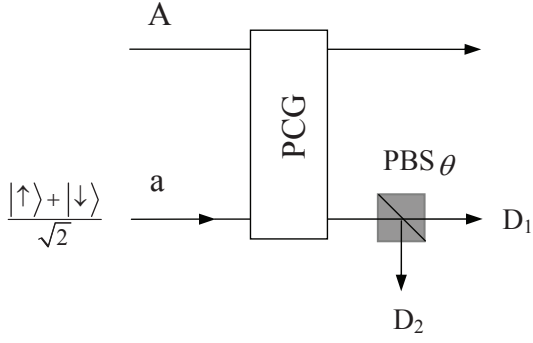}
\caption{The schematic diagram of the present entanglement
concentration protocol for electron-spin systems in a partially
entangled pure state. One of the parties in quantum communication,
say Alice performs some local operations on her electron $A$ in the
system and an additional electron $a$ in the state $\vert
+\rangle=\frac{1}{\sqrt{2}}(\vert \uparrow\rangle + \vert \downarrow
\rangle)$. PBS$_\theta$ represents a PBS whose axial direction is
placed at the angle $\theta$ along the incidence
electron.}\label{fig2}
\end{center}
\end{figure}

With the PCG shown in Fig.\ref{fig1_QND},  the principle of our ECP
for nonlocal two-electron systems in a less-entangled pure state  is
shown in Fig.\ref{fig2}. Suppose that the partially entangled pure
state for two-electron systems after they suffer from the
decoherence coming from the transmission or the  storage in a noisy
environment is
\begin{eqnarray}
|\Phi_1\rangle_{AB}=\alpha|\uparrow\rangle_{A}|\uparrow\rangle_{B} +
\beta|\downarrow\rangle_{A}|\downarrow\rangle_{B},\label{originalstate1}
\end{eqnarray}
where the subscripts $A$ and $B$ represent the two electrons shared
by two remote parties in quantum communication, say Alice and Bob.
$\alpha$ and $\beta$ are two real numbers and satisfy the relation
\begin{eqnarray}
|\alpha|^{2}+|\beta|^{2}=1.
\end{eqnarray}
Alice and Bob know these two parameters before they distill a subset
of maximally entangled electron pairs from a set of nonlocal
electron pairs in the state $|\Phi_1\rangle_{AB}$, as the same as
the ECPs with entanglement swapping and a collective unitary
evolution \cite{swapping1,swapping2}. In fact, it is not difficult
to obtain this information about the partially entangled pure state
by measuring some samples in a practical quantum communication.

For distilling a subset of maximally entangled electron pairs from a
set of pairs in a partially entangled pure state
$|\Phi_1\rangle_{AB}$, Alice prepares an additional electron $a$ in
the spin state $\vert \Phi\rangle_{a}=\frac{1}{\sqrt{2}}(\vert
\uparrow\rangle + \vert \downarrow\rangle)$ for each nonlocal
two-electron system $AB$ and then performs a parity-check
measurement on her electrons $A$ and $a$. If she obtains an even
parity, the three-electron system $ABa$ is in the state
\begin{eqnarray}
\vert\Psi_e\rangle_{ABa}=\alpha|\uparrow\rangle_{A}|\uparrow\rangle_{B}\vert
\uparrow\rangle_{a} +
\beta|\downarrow\rangle_{A}|\downarrow\rangle_{B}\vert
\downarrow\rangle_{a}.
\end{eqnarray}
If she obtains an odd parity, the system is in the state
\begin{eqnarray}
\vert\Psi_o\rangle_{ABa}=\alpha|\uparrow\rangle_{A}|\uparrow\rangle_{B}\vert
\downarrow\rangle_{a} +
\beta|\downarrow\rangle_{A}|\downarrow\rangle_{B}\vert
\uparrow\rangle_{a},
\end{eqnarray}
and Alice can transform it into the state $\vert\Psi_e\rangle_{ABa}$
by performing a bit-flip operation $\sigma_x=|\uparrow\rangle\langle
\downarrow| + |\downarrow\rangle\langle \uparrow|$ on the electron
$a$. That is, we need only describe the principle of the present ECP
when Alice and Bob obtain their electron systems in the state
$\vert\Psi_e\rangle_{ABa}$ below.

We can rewrite the state $\vert\Psi_e\rangle_{ABa}$ under the
orthogonal basis $\{ \vert \varphi\rangle_{a}=\alpha \vert
\uparrow\rangle - \beta \vert \downarrow\rangle,  \vert
\varphi^\bot\rangle_{a}=\beta \vert \uparrow\rangle + \alpha\vert
\downarrow\rangle\}$, i.e.,
\begin{eqnarray}
\vert\Psi_e\rangle_{ABa} &=& (\alpha^2 \vert
\uparrow\rangle_A|\uparrow\rangle_B - \beta^2 \vert
\downarrow\rangle_A|\downarrow\rangle_B)\vert \varphi\rangle_a \nonumber\\
&+&  \sqrt{2}\alpha\beta \cdot \frac{\vert
\uparrow\rangle_A|\uparrow\rangle_B + \vert
\downarrow\rangle_A|\downarrow\rangle_B}{\sqrt{2}}\vert
\varphi^\bot\rangle_{a}.
\end{eqnarray}
Alice can use a PBS$_\theta$, whose axial direction is placed at the
angle $\theta$ along the incidence electron, and two detectors to
complete the projection measurement on the additional electron $a$
with the basis $\{\vert \varphi\rangle_{a}, \vert
\varphi^\bot\rangle_{a}\}$, shown in Fig.\ref{fig2}. Here
$cos\theta=\alpha$ and $sin\theta=-\beta$. If Alice obtains the
state $\vert \varphi^\bot\rangle_{a}$ when she measures the
additional electron $a$, the electron pair $AB$ is in the maximally
entangled state $\vert\phi^+\rangle_{AB}=\frac{1}{\sqrt{2}}(\vert
\uparrow\uparrow\rangle + \vert \downarrow\downarrow\rangle)_{AB}$,
which takes place with the probability of $2\alpha^2\beta^2$. If
Alice obtains the state $\vert \varphi\rangle_{a}$, the electron
pair $AB$ is in another partially entangled pure state
\begin{eqnarray}
\vert\Phi_2\rangle_{AB} &=& \frac{1}{\sqrt{\alpha^4 + \beta^4}}(
 \alpha^2  \vert \uparrow\rangle_A|\uparrow\rangle_B -
 \beta^2  \vert
\downarrow\rangle_A|\downarrow\rangle_B),
\end{eqnarray}
which takes place with the probability of $\alpha^4 +
\beta^4=1-2\alpha^2\beta^2$.

It is obvious that the less-entangled pure state
$\vert\Phi_2\rangle_{AB}$ has the same form as the state
$\vert\Phi_1\rangle_{AB}$ shown in Eq.(\ref{originalstate1}). We
need only replace $\alpha$ and $\beta$ with $\alpha'\equiv
\frac{\alpha^2}{\sqrt{\alpha^4 +  \beta^4}}$ and $\beta' \equiv
\frac{\beta^2}{\sqrt{\alpha^4 +  \beta^4}}$, respectively. That is,
Alice and Bob can distill the maximally entangled state
$\vert\phi^+\rangle_{AB}$ from the state $\vert\Phi_2\rangle_{AB}$
with the probability of $2(\alpha^4 + \beta^4)\alpha'^2\beta'^2$ by
adding another additional electron $a'$ and a parity-check
measurement. Moreover, they can distill the electron pairs in the
maximally entangled state $\vert\phi^+\rangle_{AB}$ from the
less-entangled systems in the next round yet. That is, by iterating
the entanglement concentration process $n$ times, the total success
probability of this ECP is
\begin{eqnarray}
P_n &=& 2[\alpha^2\beta^2 + \frac{\alpha^4\beta^4}{\alpha^4 +
\beta^4} + \frac{\alpha^8\beta^8}{(\alpha^4 +
\beta^4)(\alpha^8 + \beta^8)} \nonumber\\
&+& \frac{\alpha^{16}\beta^{16}}{(\alpha^4 + \beta^4)(\alpha^8 +
\beta^8)(\alpha^{16}+\beta^{16})} +
\cdots\nonumber\\
&+& \frac{\alpha^{2^n}\beta^{2^n}}{(\alpha^4 + \beta^4)(\alpha^8 +
\beta^8)\cdots(\alpha^{2^n}+\beta^{2^n})}]. \label{totalprobability}
\end{eqnarray}

\begin{figure}[!h]%[tpb]
\begin{center}
\includegraphics[width=8cm,angle=0]{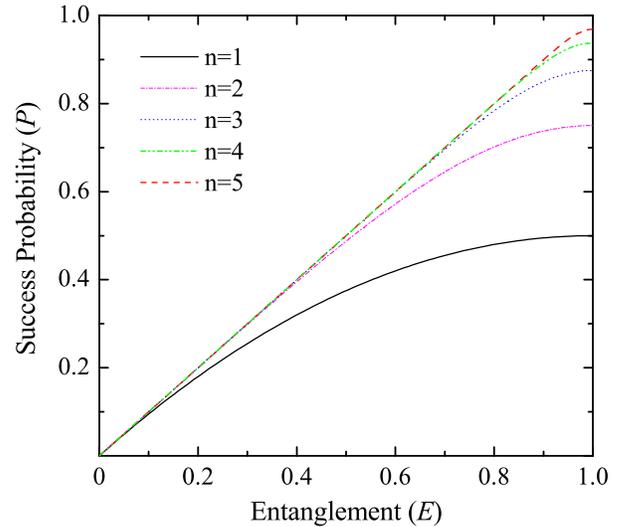}
\caption{(Color online) The relation between the success probability
of the present ECP $P$ and the entanglement of the partially
entangled state $E$ under the iteration numbers of entanglement
concentration process $n=$ 1, 2, 3, 4 and 5, respectively.
}\label{fig3}
\end{center}
\end{figure}

If $|\alpha| \leq |\beta|$, the entanglement of the  state
$|\Phi_1\rangle_{AB}=\alpha|\uparrow\rangle_{A}|\uparrow\rangle_{B}
+ \beta|\downarrow\rangle_{A}|\downarrow\rangle_{B}$ is
$E=2|\alpha|^2$ \cite{swapping1,swapping2}. The relation between the
success probability $P$ that the two parties obtain a nonlocal
two-electron system $AB$ in the maximally entangled state $\vert
\phi^+\rangle_{AB}$ from a system in the partially entangled state
$|\Phi_1\rangle_{AB}$ and the entanglement $E$ is shown in
Fig.\ref{fig3}. When $E<0.4$, Alice and Bob need only perform the
entanglement concentration process twice ($n=2$) for obtaining the
success probability $P$ nearly equivalent to the entanglement $E$.
When $0.4<E<0.72$ ($0.72<E<0.88$), they should perform the process 3
(4) times for obtaining an optimal success probability. From
Fig.\ref{fig3}, one can see that  5  times for the iteration of the
entanglement concentration process is usually enough to obtain an
optimal success probability.

\section{Optimal multipartite entanglement concentration of N-electron spin states}

It is straightforward to generalize our ECP for reconstructing
maximally entangled $N$-electron GHZ states from partially entangled
GHZ-class states.  Suppose the partially entangled $N$-electron
GHZ-class states are described as follows:
\begin{eqnarray}
|\Phi_N\rangle=\alpha|\uparrow\uparrow\cdot\cdot\cdot
\uparrow\rangle_{AB\dots Z} +
\beta|\downarrow\downarrow\cdot\cdot\cdot \downarrow\rangle_{AB\dots
Z},
\end{eqnarray}
where $|\alpha|^{2}+|\beta|^{2}=1$. The subscript $A$, $B$, $\dots$,
and $Z$ represent the electrons hold by Alice, Bob, $\dots$, and
Zach, respectively. For obtaining a subset of nonlocal $N$-electron
systems in a maximally entangled state, Alice prepares an additional
electron $a''$ in the state $\frac{1}{\sqrt{2}}(\vert \uparrow
\rangle + \vert \downarrow\rangle)$ and she performs a parity-check
measurement on her electron $A$ and the additional electron $a''$.
If she obtains an even parity, the ($N+1$)-electron system is in the
state
\begin{eqnarray}
|\Phi_{N+1}\rangle_e &=& \alpha|\uparrow\rangle_A\vert
\uparrow\rangle_{a''}\vert\uparrow\cdot\cdot\cdot
\uparrow\rangle_{B\dots Z} \nonumber\\
 &+&
\beta|\downarrow\rangle_A\vert
\downarrow\rangle_{a''}\vert\downarrow\cdot\cdot\cdot
\downarrow\rangle_{B\dots Z}.
\end{eqnarray}
If Alice obtains an odd parity, the system is in the state
\begin{eqnarray}
|\Phi_{N+1}\rangle_o &=& \alpha|\uparrow\rangle_A\vert
\downarrow\rangle_{a''}\vert\uparrow\cdot\cdot\cdot
\uparrow\rangle_{B\dots Z} \nonumber\\
 &+&
\beta|\downarrow\rangle_A\vert
\uparrow\rangle_{a''}\vert\downarrow\cdot\cdot\cdot
\downarrow\rangle_{B\dots Z}.
\end{eqnarray}
Obviously, the state $|\Phi_{N+1}\rangle_o$ can be transformed into
the state $|\Phi_{N+1}\rangle_e$ with a bit-flip operation
$\sigma_x$ on the additional electron $a''$. By projecting the state
of the additional electron $a''$ into the orthogonal basis $\{\vert
\varphi\rangle_{a''}=\alpha \vert \uparrow\rangle - \beta \vert
\downarrow\rangle, \vert \varphi^\bot\rangle_{a''}=\beta \vert
\uparrow\rangle + \alpha\vert \downarrow\rangle\}$, the $N$ parties
obtain the maximally entangled state $\vert\phi^+\rangle
=\frac{1}{\sqrt{2}}(\vert \uparrow\uparrow \cdot\cdot\cdot
\uparrow\rangle + \vert \downarrow\downarrow \cdot\cdot\cdot
\downarrow\rangle)_{AB\dots Z}$ directly if Alice obtains the state
of the additional electron $\vert \varphi^\bot\rangle_{a''}$, which
takes place with the probability of $2\alpha^2\beta^2$. If Alice
obtains the state $\vert \varphi\rangle_{a''}$, the parties  obtain
a partially less-entangled pure state
\begin{eqnarray}
|\Phi'_N\rangle=\frac{1}{\sqrt{\alpha^4 +
\beta^4}}(\alpha^2|\uparrow\uparrow\cdot\cdot\cdot
\uparrow\rangle_{AB\dots Z} +
\beta^2|\downarrow\downarrow\cdot\cdot\cdot
\downarrow\rangle_{AB\dots Z}).\nonumber\\
\end{eqnarray}
The parties can also distill some maximally entangled state from
this partially less-entangled state, similar to the case with
nonlocal two-electron systems in a partially entangled pure state.
That is, this ECP works for $N$-electron systems in a partially
entangled pure state yet.

\section{Discussion and summary}

Let us compare the efficiency of the present ECP for electron
systems with those in others
\cite{Bennett2,swapping1,swapping2,shengplaec}. Of course, there are
some differences between the present ECP and others. The present ECP
requires that the parties know the information about the initial
state of the nonlocal $N$-electron systems, as the same as those in
Refs. \cite{swapping1,swapping2}. However, these in Refs.
\cite{Bennett2,shengplaec} do not require the parties to know the
information. In essence, all existing ECPs
\cite{Bennett2,swapping1,swapping2,shengplaec} for electron systems
are based on the Schmidt projection method in which the parties
exploit the combination of a pair of systems with the same parameter
to obtain a system in a maximally entangled state with an average
success probability of $\alpha^2\beta^2$. The ECP in Ref.
\cite{swapping2} exploits an additional collective unitary evolution
on one qubit in the system and an additional qubit to improve the
success probability to be $\alpha^2$. The ECP for electron systems
in Ref. \cite{shengplaec} simplified the implementation by
sacrificing the efficiency, compared with those in Refs.
\cite{swapping1,swapping2}. However, the present ECP distills an
$N$-electron system from a system in a partially entangled pure
state and an additional electron, not a pair of $N$-electron
systems, which is far  different from all existing ECPs
\cite{Bennett2,swapping1,swapping2,shengplaec}. Moreover, the
success probability of the present ECP is  $2\alpha^2$ ($|\alpha|$
$\leq |\beta|$) which is just the entanglement of the partially
entangled pure state. That is, the success probability of the
present ECP is the theoretical limit of ECPs for a partially
entangled pure state. It is an optimal one.

%(Color online)

\begin{figure}[!h]%[tpb]
\begin{center}
\includegraphics[width=8cm,angle=0]{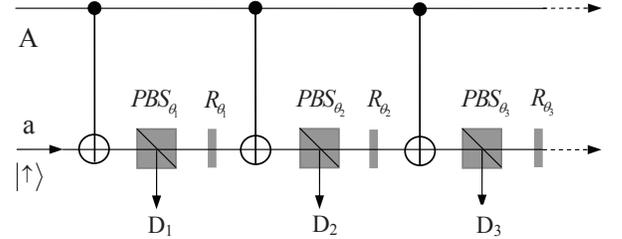}
\caption{The circuit for the present ECP with CNOT gates.
}\label{fig4}
\end{center}
\end{figure}

We present our ECP with PCGs. In fact, it works with CNOT gates
efficiently. The circuit is shown in Fig.\ref{fig4}.
$PBS_{\theta_1}$ transmits the spin state of the additional electron
$\alpha \vert\uparrow\rangle_a - \beta \vert \downarrow\rangle_a$
and reflects the spin state $\beta \vert\uparrow\rangle + \alpha
\vert \downarrow\rangle$. $R_{\theta_1}$ is used to rotate the spin
state of the additional electron $\alpha \vert\uparrow\rangle_a -
\beta \vert \downarrow\rangle_a$ to be $\vert \uparrow\rangle_a$.
$PBS_{\theta_2}$ transmits the spin state $\frac{1}{\sqrt{\alpha^4
+\beta^4 }}(\alpha^2 \vert\uparrow\rangle - \beta^2 \vert
\downarrow\rangle)_a$ and reflects the spin state
$\frac{1}{\sqrt{\alpha^4 +\beta^4 }}(\beta^2 \vert\uparrow\rangle +
\alpha^2 \vert \downarrow\rangle)_a$. $R_{\theta_2}$ is used to
rotate the spin state of the additional electron
$\frac{1}{\sqrt{\alpha^4 +\beta^4 }}(\beta^2 \vert\uparrow\rangle +
\alpha^2 \vert \downarrow\rangle)_a$ to be $\vert
\uparrow\rangle_a$. $PBS_{\theta_n}$ transmits the spin state
$\frac{1}{\sqrt{\alpha^{2^{n}} + \beta^{2^{n}} }}(\alpha^{2^{n-1}}
\vert\uparrow\rangle - \beta^{2^{n-1}} \vert \downarrow\rangle)_a$
and reflects the spin state $\frac{1}{\sqrt{\alpha^{2^{n}}
+\beta^{2^{n}} }}(\beta^{2^{n-1}} \vert\uparrow\rangle +
\alpha^{2^{n-1}} \vert \downarrow\rangle)_a$. $R_{\theta_n}$ is used
to rotate the spin state of the additional electron
 $\frac{1}{\sqrt{\alpha^{2^{n}}
+\beta^{2^{n}} }}(\beta^{2^{n-1}} \vert\uparrow\rangle +
\alpha^{2^{n-1}} \vert \downarrow\rangle)_a$ to be $\vert
\uparrow\rangle_a$. For the partially entangled pure state $\vert
\Phi_1\rangle_{AB}=\alpha \vert \uparrow\rangle_A\vert
\uparrow\rangle_B + \beta \vert \downarrow\rangle_A\vert
\downarrow\rangle_B$, Alice performs a CNOT operation on her
electron $A$ and the additional electron $a$ in the state $\vert
\uparrow\rangle_a$, which makes the state of the three-electron
system $ABa$ become
$\vert\Psi_e\rangle_{ABa}=\alpha|\uparrow\rangle_{A}|\uparrow\rangle_{B}\vert
\uparrow\rangle_{a} +
\beta|\downarrow\rangle_{A}|\downarrow\rangle_{B}\vert
\downarrow\rangle_{a}$. The $PBS_{\theta_1}$ will project the state
of the additional electron $a$ into the orthogonal basis $\{\vert
\varphi\rangle_{a}=\alpha \vert \uparrow\rangle - \beta \vert
\downarrow\rangle, \vert \varphi^\bot\rangle_{a}=\beta \vert
\uparrow\rangle + \alpha\vert \downarrow\rangle\}$. When the
additional electron is filtered out from the output $D_1$, the
two-electron system $AB$ is in the maximally entangled state $\vert
\phi^+\rangle_{AB} =\frac{1}{\sqrt{2}}(\vert \uparrow\uparrow\rangle
+ \vert \downarrow\downarrow\rangle)_{AB}$. Otherwise, the
additional electron will be recovered to be the initial state $\vert
\uparrow\rangle_a$ and enter into the next round of entanglement
concentration. When the additional electron is detected at the
outputs $D_1$, $D_2$, $\cdots$, or $D_n$, the two-electron system
$AB$ will be in the maximally entangled state.

In summary, we have proposed  an optimal ECP for nonlocal
$N$-electron systems in a  partially entangled pure state, resorting
to charge detection and the projection measurements on additional
electrons. One of the $N$ parties, say Alice exploits the PCG based
on charge detection to extend the partially entangled $N$-electron
system to an $(N+1)$-electron system first and then she projects the
additional electron into an orthogonal basis. By detecting the
output of the additional electron from a PBS, the $N$-parties in
quantum communication can divide their $N$-electron systems into two
groups. One is in the maximally entangled state. The other is in
another partially entangled state with less entanglement, which is
just the resource for the entanglement concentration in the next
round. By iterating the entanglement concentration process several
times, the $N$ parties can obtain a subset of $N$-electron systems
in the maximally entangled state with the maximal success
probability which is just equivalent to the entanglement of the
partially entangled state. Compared with other ECPs
\cite{Bennett2,swapping1,swapping2,shengplaec}, the present ECP has
the optimal success probability, the theoretical limit, without
resorting to a collective unitary evolution.\\

This work is supported by the National Natural Science Foundation of
China under Grant Nos. 10974020 and 11174039,  NCET, and the
Fundamental Research Funds for the Central Universities.

\end{document}